\documentclass[preprint,1p,times]{elsarticle}

\usepackage{amssymb}

\journal{Physics Letters A}

\begin{document}

\begin{frontmatter}



\title{Mass of a lattice polaron from an extended Holstein model using the
    Yukawa potential}


\author{B.Ya.Yavidov, Sh.S.Djumanov and S.Dzhumanov}

\address{Institute of Nuclear Physics, 100214 Ulughbek, Tashkent, Uzbekistan}

\begin{abstract}
Renormalization of the mass of an electron is studied within the framework of the Extended Holstein model at strong coupling regime and nonadiabatic limit. In order to take into account an effect of screening of an electron-phonon interaction on a polaron it is assumed that the electron-phonon interaction potential has the Yukawa form and screening of the electron-phonon interaction is due to the presence of other electrons in a lattice. The forces are derived from the Yukawa type electron-phonon interaction potential. It is emphasized that the early considered screened force of Refs.\cite{kor-giant,spencer,hague-etal,hague-kor} is a particular case of the force deduced from the Yukawa potential and is approximately valid at large screening radiuses compared to the distances under consideration. The Extended Holstein polaron with the Yukawa type potential is found to be a more mobile than polaron studied in early works at the same screening regime.
\end{abstract}


\end{frontmatter}

A model of a polaron with a long-range "density-displacement" type force was introduced by Alexandrov and Kornilovitch in Ref.\cite{alekor}.
The model by itself represents an extension of the Fr\"{o}hlich polaron model \cite{froh} to a discrete ionic crystal lattice or extension of the
Holstein polaron model \cite{hol} to a case when an electron interacts with many ions of a lattice with longer ranged electron-phonon interaction.
Subsequently, the model was named as the extended Holstein model (EHM) \cite{flw}. The model was introduced in order to mimic  $high-T_{c}$ cuprates,
where the in-plane ($CuO_{2}$) carriers are strongly coupled to the $c$-axis polarized vibrations of the $apical$ oxygen ions \cite{timusk}.
In the last decade the model was successfully applied to cuprates
\cite{kor-ctqmc,kor-giant,kor-ground,flw,bt,trg,b-trg,asa-kor-pha,asa-kor-jpcm,pcf-jpcm,cfmp-prb,hohen,asa-ya,spencer,hague-etal,
kor-jpcm,hague-sam,hague-kor,yav-jetp,yav-physb} as well as to semiconducting polymers \cite{stojan,meisel}.
Kornilovitch in Ref. \cite{kor-ctqmc} studied the ground state energy, effective mass and polaron spectrum with the help of continuous-time Quantum
Monte Carlo algorithm. An anisotropy of polaron's mass due to electron-phonon interaction, ground-state dispersion and density of states of a EHM polaron
 were studied in Ref.\cite{kor-giant} and Ref.\cite{kor-ground}, respectively. Fehske, Loos and Wellein \cite{flw} investigated the electron-lattice
 correlations, single-particle spectral function and optical conductivity of a polaron within the EHM in the strong and weak coupling regimes by means of
  an  exact Lancroz diagonalization method. Other properties of EHM, such as the ground state spectral weight, the average kinetic energy and the mean
  number of phonons were studied in \cite{pcf-jpcm,cfmp-prb,hohen} by means of the variational and Quantum Monte Carlo simulation approaches.
  The work \cite{asa-ya} extended the EHM to the adiabatic limit. The effect of the different type polarized vibrations of ions and the arrangement of
  the ions  on mass of a polaron was studied in Refs.\cite{trg,yav-jetp,yav-physb}. The EHM with screened electron-phonon interaction was discussed in
  Refs.\cite{kor-giant,spencer,hague-etal,hague-kor}.
At the same time polarons were experimentally recognized as quasiparticles in the novel materials, in particular, in the superconducting
cuprates and colossal magnetoresistance manganites \cite{kim,sharma,haga,mihai,egami-91,jorgenson,zhong,bar-bishop,flack,calvani,reitschel,egami-96,zhao,temprano,lanzara,shen,rosch}.
For details on polaronic effects in cuprates and other novel materials we refer a reader to the review papers and books (see for example Ref.
 \cite{salje-asa} and Ref.\cite{devr-asa}).
In this paper a particular question of coupled electron-phonon system will be considered within the EHM. Namely, it is an influence of screened
electron-phonon interaction on mass of a polaron. In contrast to Refs.\cite{kor-giant,spencer,hague-etal,hague-kor} an explicit form of electron-phonon
interaction forces will be derived associated with the different type of polarized vibrations and renormalized mass of a EHM polaron at different values
of screening radius will be calculated. We will see that an effect of screening is more pronounced at small values of the screening radius and discuss
possible consequences of early applied approximation.
We consider an electron performing hopping motion on a lower chain consisting of the static sites, but interacting with all ions of an upper chain via
 a long-range density-displacement type force, as shown in Fig.1. So, the motion of an electron is always one-dimensional, but a vibration of the upper
 chain's ions is isotropic and two-dimensional one.
\begin{figure}[tbp]
\begin{center}
\includegraphics[angle=-0,width=0.75\textwidth]{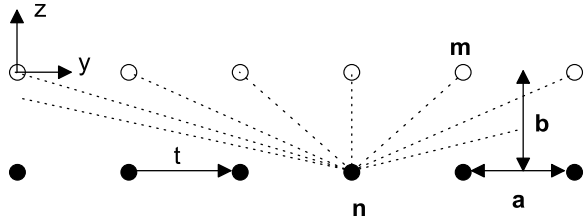} \vskip -0.5mm
\end{center}
\caption{An electron hops on a lower chain and
interacts with the ions vibrations of an upper infinite chain via a density-displacement type force $f_{{\bf m},\alpha}(\bf n)$.
The distances between the chains ($|{\bf b}|$) and between the ions ($|{\bf a}|$) are assumed equal to 1. Dotted lines represents an interaction of an electron on site ${\bf n}$ with the ions of the upper chain.}
\end{figure}
The Hamiltonian of the model  is
\begin{equation}\label{1}
H=H_{e}+H_{ph}+H_{e-ph}
\end{equation}
where
\begin{equation}\label{2}
H_{e}=-t \sum_{\bf n}(c^{\dagger}_{\bf n}c_{\bf n+a}
+H.c.)
\end{equation}
is the electron hopping energy,
\begin{equation}\label{3}
H_{ph}=\sum_{{\bf m},\alpha}\left(-\frac{\hbar^2\partial^2}{2M\partial u^{2}_{{\bf m},\alpha}}+\frac{M\omega^2u^{2}_{{\bf m},\alpha}}{2}\right)
\end{equation}
is the Hamiltonian of the vibrating ions,
\begin{equation}\label{4}
H_{e-ph}=\sum_{{\bf n,m},\alpha}f_{{\bf m},\alpha}({\bf n})\cdot u_{{\bf m},\alpha}c^{\dagger}_{\bf n}c_{\bf n}
\end{equation}
describes interaction between the electron that belongs to a lower chain and the ions of an upper chain. Here $t$ is the nearest neighbor hopping integral, $c^{\dagger}_{\bf n}$($c_{\bf n}$) is a creation (destruction) operator of an electron on a cite $\bf n$, $u_{{\bf m},\alpha}$ is the $\alpha=y,z$- polarized displacement of the {\bf m}-th ion and $f_{{\bf m},\alpha}({\bf n})$ is an interacting density-displacement type
force between an electron on a site {\bf n} and the $\alpha$ polarized vibration of the {\bf m}-th ion. $M$ is the mass of the vibrating ions
and $\omega$ is their frequency. There is no doubt that an explicit analytical form of the force $f_{\bf {m},\alpha}({\bf n})$ is one of crucial aspects determining polaron parameters. Of cause, it depends on structural elements that located on sites $\bf m$. Whether the structural elements are neutral, charged (positively or negatively) or dipoles (electrical or magnet) this force may have different origin and may lead to different polaronic states. As in Refs.\cite{kor-giant,spencer,hague-etal,hague-kor} here it is assumed that the structural elements are electrically charged (positively or negatively) and thus the force has Coulombic nature. The early studies in the EHM were performed with the unscreened electron-phonon interaction where the force is deduced from pure Coulomb potential $\approx const/r$ which is for our discrete lattice is written as $\approx const/\sqrt{|{\bf n}-{\bf m}|^2+b^2}$. The distance along the chain $|{\bf n}-{\bf m}|$ is measured in the units of a lattice constant $|{\bf a}|=1$. The distance between the chains is $|{\bf b}|=1$ too. Detailed derivation of the unscreened force can be found in Ref.\cite{spencer}. The EHM was studied in Refs.\cite{kor-giant,spencer,hague-etal,hague-kor} with screened forces in which
the screening effect due to the presence of other electrons in the lattice. However, an explicit derivation of the screened force was not presented. In general, at present there is no the exact analytical expression for a screened electron-ion force. Commonly used formulas for screened forces are obtained under some approximations. Here we consider a more general form of electron-phonon interaction force due to direct Coulomb forces of an electron in the lower chain with ions of the upper chain. Namely, we approximate an electron-ion interaction potential as the Yukawa potential: $\approx const\exp[-r/R]/r$, where $R$ is the screening radius and $r$ is the position radius. Such a type of approximation is more suitable to real systems, in particular, in the case of cuprates. Indeed, cuprates change their properties upon doping from insulating state to metallic one. In such circumstances, choice of Yukawa potential seems to be appropriate since one has to consider different doping regimes. In optimally and overdoped regimes, cuprates are believed to be in a metallic state and one expects the form of electron-ion potential would be $(Ze^2/r)\exp{[-r/\lambda_{TF}]}$, where $\lambda_{TF}=(E_F/2\pi e^2n_0)^{1/2}$ is the Thomas-Fermi screening radius, $E_F$ is the Fermi energy, $n_0$ is an equilibrium charge density. In the opposite underdoped regime cuprates are likely in a semiconducting state and thus one can use Debye approximation for the screened electron-ion potential $(Ze^2/r)\exp{[-r/\lambda_D]}$, where $\lambda_D=(\varepsilon_0 k_BT/4\pi e^2n_0)^{1/2}$ is the Debye screening radius, $T$ is absolute temperature, $\varepsilon_0$ static dielectric constant of cuprates and $k_B$ is the Boltzmanm constant. As in the both regimes electron-ion potentials have exponential terms our choice seems to be a more realistic. Then a discrete form of the electron-ion potential is written as:
\begin{eqnarray}
  U_{{\bf m}}({\bf n}) &=& \frac{\kappa}{(|{\bf n}-{\bf m}|^2+b^2)^{1/2}}\times \\
\nonumber   &\times& \exp\left[-\frac{\sqrt{|{\bf n}-{\bf m}|^2+b^2}}{R}\right]
\end{eqnarray}
where $\kappa$ is some coefficient and $R$ is measured in units of $|\bf a|$. From the potential Eq.(5) one obtain an analytical expressions for the $z$- and $y$- type components of the screened electron-phonon forces:
\begin{equation}\label{6}
f_{{\bf m},y}({\bf n})=\frac{\kappa |{\bf n}-{\bf m}|}{(|{\bf n}-{\bf m}|^2+b^2)^{3/2}}\times\left(1+\frac{\sqrt{|{\bf n}-{\bf m}|^2+b^2}}{R}\right)\times\exp\left[-\frac{\sqrt{|{\bf n}-{\bf m}|^2+b^2}}{R}\right]
\end{equation}
and
\begin{equation}\label{7}
f_{{\bf m},z}({\bf n})=\frac{\kappa b}{(|{\bf n}-{\bf m}|^2+b^2)^{3/2}}\times\left(1+\frac{\sqrt{|{\bf n}-{\bf m}|^2+b^2}}{R}\right)\times\exp\left[-\frac{\sqrt{|{\bf n}-{\bf m}|^2+b^2}}{R}\right]
\end{equation}
A surprising point of a such determination of the forces from Eq.(5) is that $y$- component $f_{{\bf m},y}({\bf n})$ has no effect to an electron when ${\bf m}={\bf n}$. This is not the case if ions of the upper chain are shifted along $y$- direction by some distance as in \cite{bt,trg,b-trg}. As one can see from Eq.(7) the screened electron-phonon interaction force used in Refs.\cite{kor-giant,spencer,hague-etal,hague-kor} (see for example Eq.(16) of Ref.\cite{spencer}) is a particular case of the more general type of force induced by the Yukawa type potential Eq.(5). Indeed, if one assumes validity of the conditions $|{\bf b}|\ll |{\bf n}-{\bf m}|\ll R$, our Eq.(7) reduces to Eq.(16) of Ref.\cite{spencer}. Thus the early studied screened electron-phonon interaction force is represents a particular case of Eq.(7) at large screening radius $R$ compared to the distance from an electron to the distant ions $|{\bf n}-{\bf m}|$. Unscreened force of Ref.\cite{alekor} ($z$- component) and Ref.\cite{bt} may be considered as the particular cases of our force at $R=\infty$ and $R=1$, respectively.
\begin{figure}[tbp]
\begin{center}
\includegraphics[angle=-0,width=0.75\textwidth]{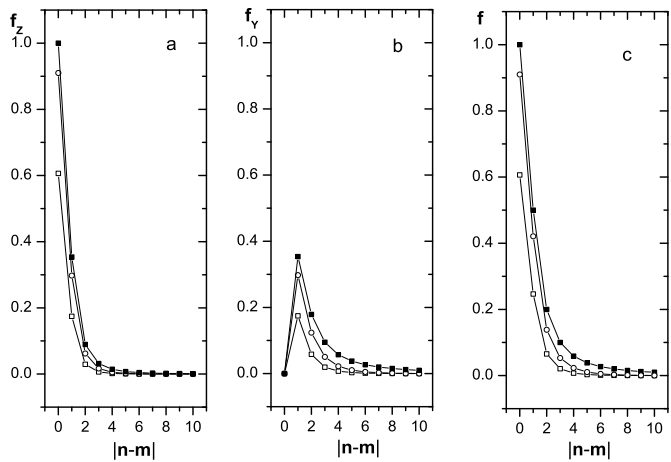} \vskip -0.5mm
\end{center}
\caption{The values of electron-phonon interaction forces as a function of $|{\bf n}-{\bf m}|$. Filled squares ($\blacksquare$), open squares ($\square$) and open circles ($\circ$) correspond to unscreened, screened according to Refs.\cite{kor-giant,spencer,hague-etal,hague-kor} and to our case Eq.(5) and Eq.(6), respectively. Forces are in units of $\kappa$ and screened forces are calculated at $R=2$.}
\end{figure}
\begin{figure}[tbp]
\begin{center}
\includegraphics[angle=-0,width=1.0\textwidth]{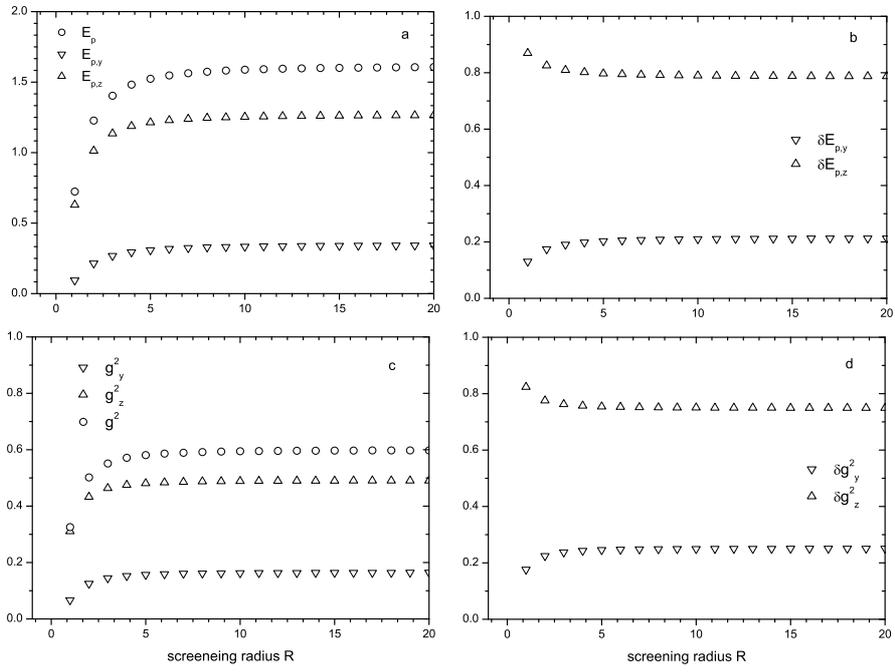} \vskip -0.5mm
\end{center}
\caption{Plot of EHM polaron parameters as a function of screening radius $R$: (a) Polaron shift(in units of $\kappa^2/2M\omega^2$), (b) contributions to a full polaron shift, (c) polaron band narrowing factors (in units of $\kappa^2/2M\hbar\omega^3$), (d) contributions to a full polaron band narrowing factor. Symbols $\bigtriangleup$ and $\bigtriangledown$ represents contributions coming from $z$- and $y$- polarized vibrations, respectively and open circles symbols $\circ$ are the overall effect of both contributions.}
\end{figure}
\begin{figure}[tbp]
\begin{center}
\includegraphics[angle=-0,width=0.75\textwidth]{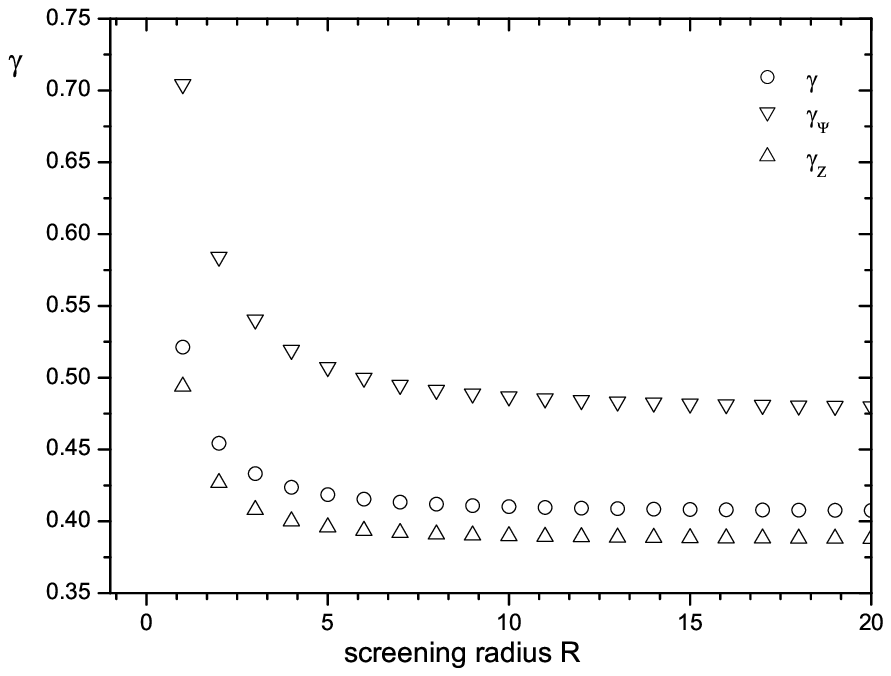} \vskip -0.5mm
\end{center}
\caption{Plot of $\gamma$ as a function of screening radius $R$. Symbols $\bigtriangleup$ and $\bigtriangledown$ represents contributions coming from $z$- and $y$- polarized vibrations, respectively and open circles symbols $\circ$ are the overall effect of both contributions.}
\end{figure}
The comparison of the unscreened and screened forces are presented in Fig.2. The dependencies of $z$- and $y$- components of the forces on $|{\bf n}-{\bf m}|$ are given in Fig.2(a) and Fig.2(b) respectively. While the dependence of the full force $f_{{\bf m}}({\bf n})=\sqrt{f_{{\bf m},z}^2({\bf n})+f_{{\bf m},y}^2({\bf n})}$  on $|{\bf n}-{\bf m}|$ is plotted in Fig.2(c). One can see that at the large distances $|{\bf n}-{\bf m}|$ the difference between forces is small. However, at small $|{\bf n}-{\bf m}|$ the forces are strongly deviate from each other. Here the short distances are of considerable interest as: (i) doping of cuprates reduces screening radius $R$ and, (ii) therefor, the polaronic effects come only from neighboring ions. The importance of nearest neighbor ions  and their arrangement in determining polaron parameters was already recognized \cite{bt,trg}.
Let's discuss an effect of screening radius on certain polaron parameters such as polaron shift $E_p$ and band narrowing factor  $g^2$ at strong coupling regime $\lambda=E_p/zt> 1$ (where $z$ is the crystal lattice coordination number) and nonadiabatic limit $t/\hbar\omega<1$. At strong coupling regime  and nonadiabatic limit one can use an analytical method based on the extended (or nonlocal) Lang-Firsov transformation \cite{alekor,flw,bt} and subsequently perturbation theory with respect to the parameter $1/\lambda$. Here we present only the resulting analytical expressions for a mass of a polaron in the EHM. According to Ref.\cite{alekor}, the mass of the EHM polaron with a single dispersionless phonon mode is given by (in units of the bare band mass) $m_p=\exp{[g^2]}$, where $g^2=g^2_z+g^2_y$ is the full band narrowing factor ,
\begin{equation}\label{8}
    g^2_\alpha=\frac{1}{2M\hbar\omega^3}\sum_{{\bf m}}[f^2_{{\bf m},\alpha}({\bf 0})-f_{{\bf m},\alpha}({\bf 0})f_{{\bf m},\alpha}({\bf 1})]
\end{equation}
is the band narrowing factor  due to the only $\alpha$- polarized vibrations. The EHM polaron mass can be expressed in terms of electron-phonon coupling constant as
\begin{equation}\label{9}
    m_p=\exp{[\gamma (E_p/\hbar\omega)]}=\exp[2\lambda\gamma t/\hbar\omega],
\end{equation}
where
\begin{equation}\label{10}
    E_p=E_{p,z}+E_{p,y}; E_{p,\alpha}=\frac{1}{2M\omega^2}\sum_{{\bf m}}f^2_{{\bf m},\alpha}({\bf 0})
\end{equation}
and
\begin{equation}\label{11}
    \gamma=1-\frac{\sum_{{\bf m},\alpha}f_{{\bf m},\alpha}({\bf 0})\cdot f_{{\bf m},\alpha}({\bf 1})}{\sum_{{\bf m},\alpha} f_{{\bf m},\alpha}^2({\bf 0})}.
\end{equation}
As is seen from Eqs.(6)-(11), all polaron parameters affected by $R$ as it enters to the all expressions. We are interested in how the polaron parameters ($E_p$, $g^2$ and $\gamma$) change with $R$. In order to elucidate this we first consider screening effect when one has the only $z$- or $y$- polarized vibrations. Then we take into account both contributions coming from each polarized vibrations. Polaron shift due to $z$- ($y$-) polarized vibrations of ions of an upper chain $E_{p,z}$ ($E_{p,y}$) and net polaron shift due to both polarized vibrations $E_p$ are given in Fig.3(a) as a function of screening radius $R$. In an analogy way band narrowing factor  $g^2$ is given in Fig.3(c). From Fig.3(a) and Fig.3(c) one can see that polaron shift and band narrowing factor  increase with screening radius. As  $R\rightarrow\infty$ the values of all of them $E_{p,z}, E_{p,y}, E_{p}, g^2_{z}, g^2_{y}, g^2$ approach to the limiting values. Calculation of these parameters at $R=\infty$ yield $E_{p,z}=1.27\kappa^2/2M\omega^2$, $E_{p,y}=0.34\kappa^2/2M\omega^2$, $E_p=1.61\kappa^2/2M\omega^2$, $g^2_z=0.49\kappa^2/2M\hbar\omega^3$, $g^2_y=0.16\kappa^2/2M\hbar\omega^3$ and $g^2=0.65\kappa^2/2M\hbar\omega^3$. When the screening radius $R$ is large enough compared to the lattice constant $|{\bf a}|$, the effect of screening on polaron shift and band narrowing factor is not so sensitive. However, as screening radius $R$ is decreased, the effect becomes more sensitive. At $R=1$ the full polaron shift is $\approx 45\%$ of the full polaron shift with unscreened electron-phonon interaction. The contributions to the full polaron shift coming from each type of polarized vibrations decreases as $R$ vanishes as well. $E_{p,z}$ and $E_{p,y}$ at $R=1$ make up approximately $50\%$ and $26\%$ of $E_{p,z}$ and $E_{p,y}$ with unscreened interaction. In Table I the calculated values of some polaron parameters are given. The relative contributions coming from the $z$- and $y$- polarized vibrations to the full polaron shift and band narrowing factor are calculated according formulas $\delta E_{p,\alpha}=E_{p,\alpha}/E_p$ and $\delta g^2_\alpha=g^2_\alpha/g^2$ (Fig.3(b) and (d)). As one can see from Fig.3(b) and Table I, $\delta E_{p,z}$ ($\delta E_{p,y}$) decreases (increases) with screening radius $R$ and approaches to its limiting value at $R=\infty$ which is $78\%$ ($22\%$). The same character of changing is true for band narrowing factor  $g^2$ and its components $g^2_z$ and $g^2_y$.  In all interval of $R$ main contribution to the polaronic effect is due to $z$- polarized vibrations.
\begin{table*}
\caption{\label{tab:table1} The calculated values of polaron parameters. The screening radius $R$, the polaron shifts ($E_p$, $E_{p,z}$ and $E_{p,y}$) and band narrowing factor s ($g^2$, $g^2_z$ and $g^2_y$) are given in units of $|{\bf a}|$, $\kappa^2/2M\omega^2$ and $\kappa^2/2M\hbar\omega^3$, respectively.}
\begin{tabular}{cccccccccccccc}
$R$&$E_{p,z}$&$E_{p,y}$&$E_p$&$g^2_z$&$g^2_y$&$g^2$&$\delta E_{p,z}$&$\delta E_{p,y}$&$\delta g^2_z$&$\delta g^2_y$&$\gamma_z$&$\gamma_y$&$\gamma$\\
\hline
       1&0.63&0.09&0.72&0.31&0.06&0.37&0.87&0.13&0.82&0.17&0.49&0.70&0.52\\
       2&1.01&0.21&1.22&0.43&0.12&0.55&0.83&0.17&0.77&0.23&0.42&0.58&0.45\\
       3&1.14&0.26&1.40&0.46&0.14&0.60&0.81&0.19&0.76&0.24&0.41&0.54&0.43\\
       4&1.19&0.29&1.48&0.47&0.15&0.62&0.80&0.20&0.75&0.25&0.40&0.52&0.42\\
       5&1.21&0.31&1.52&0.48&0.16&0.64&0.79&0.21&0.75&0.25&0.39&0.50&0.41\\
$\infty$&1.27&0.34&1.61&0.49&0.16&0.65&0.78&0.22&0.75&0.25&0.38&0.47&0.40\\
\end{tabular}
\end{table*}
Screened electron-phonon interaction has also an effect on parameter $\gamma$ which determines mass of a polaron Eq.(9). This effect depends on a type of electron-ion potential (or electron-ion force) and on structure of a lattice. We have considered a situation when a polaron is formed by (i) the only $z$- ($y$-) polarized vibrations of the upper chain ions and (ii) the both $z$- and $y$- polarized vibrations of the upper chain. For each situation we have calculated $\gamma_z$ ($\gamma_y$) and $\gamma$ for a lattice depicted in Fig.1 at different $R$. The results are presented in Fig.4 and in the last three column of Table I. The general tendency is that all of them decreases with screening radius $R$. The calculated masses of the EHM polarons as a function of the electron-phonon coupling constant $\lambda$ for $R=1$ (thin line), $R=2$ (short-dotted line), $R=3$ (dash-dotted line), $R=4$ (dotted line), $R=5$ (dashed line) and $R=\infty$ (solid thick line) are plotted in Fig.5 at $t/\hbar\omega=0.5$. We confirm early findings which indicate that unscreened electron-phonon interaction provides a more mobile polaron and polaron with the screened electron-phonon interction has a more renormalized mass \cite{spencer}. At the same time our study differs from the early studied works by the form of electron-phonon force. This form is derived with the help of more general form of electron-ion interaction potential which is Yukawa potential and can be microscopically derived (see for example \cite{mahan}). At the regimes when screening radius is comparable to the lattice constant which may be reached by the doping of a sample our force strongly deviates from early studied force. It lies somewhere in the middle of the totally unscreened force of Ref.\cite{alekor} and the screened force of Ref.\cite{spencer}(Fig.2). This has a serious impact on the whole range of polaron parameters, in particulary to the $\gamma$. Calculating of $\gamma_z$ with a more general form of electron-phonon interaction force Eq.(7) one finds $\gamma_z=0.49$ and $\gamma_z=0.41$ for $R=$ 1 and $R=3$, respectively. These results should be compared with $\gamma_z=0.75$ and $\gamma_z=0.53$ of Ref.\cite{spencer} for the same $R=1$ and $R=3$, respectively. In this sense use of Yukawa potential provides a more mobile polaron of EHM at any range of screening radius.
\begin{figure}[tbp]
\begin{center}
\includegraphics[angle=-0,width=0.75\textwidth]{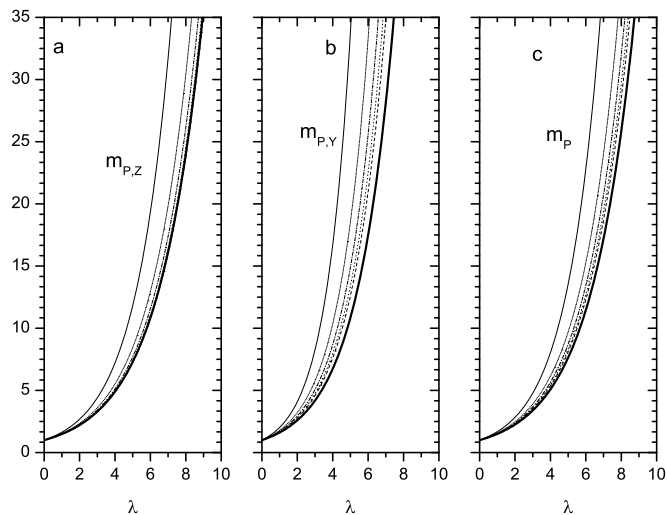} \vskip -0.5mm
\end{center}
\caption{EHM polaron mass (a) $m_{p,z}(m_{p,y})$ as a function of electron-phonon coupling constant $\lambda$ due to only $z$ ($y$)- polarized vibrations of ions and (b) full polaron mass when both polarizations contribute to a mass renormalization at different values of screening radius $R$: $R=1$- thin line, $R=2$- short-dotted line, $R=3$- dash-dotted line, $R=4$- dotted line, $R=5$- dashed line and $R=\infty$- solid thick line.   $t/\hbar\omega=0.5$.}
\end{figure}

In conclusion we have studied an extended Holstein model with the Yukawa type electron-phonon interaction potential. A more general form of the density-displacement type electron-phonon interaction force is derived for a discrete lattice. The early considered force of Ref.\cite{kor-giant,spencer,hague-etal,hague-kor} is a particular case of the force studied here. At the large values of screening radius these forces decay exponentially with $|{\bf n}-{\bf m}|$ and difference between them is small. However, at small values of the screening radius the forces have clear -cut distinction. As a consequence, the EHM polaron with the Yukawa type electron-phonon interaction potential is found to be a more mobile than a polaron of Ref.\cite{kor-giant,spencer,hague-etal,hague-kor} for the same screening radius $R$.

One of us (B.Ya.Ya.) is grateful to Dr M. Ermamatov for valuable and fruitful discussions. This work is supported by Uzbek Academy of Science, Grant No. ${\Phi}$A-${\Phi}$2-${\Phi}$070.


\end{document}